\documentclass[a4paper,11pt]{article}
\usepackage{pos}
\usepackage{url}
\usepackage{breakurl}

\title{Towards Equitable, Diverse, and Inclusive science collaborations: The Multimessenger Diversity Network}
 \ShortTitle{Multimessenger Diversity Network}

\author{The IceCube Collaboration and the Multimessenger Diversity Network\\{\normalsize \normalfont(a complete list of authors can be found at the end of the proceedings)}}

\emailAdd{ellen.bechtol@icecube.wisc.edu}
\emailAdd{jim.madsen@icecube.wisc.edu}

\abstract{The Multimessenger Diversity Network (MDN), formed in 2018, extends the basic principle of multimessenger astronomy – that working collaboratively with different approaches enhances understanding and enables previously impossible discoveries – to equity, diversity, and inclusion (EDI) in science research collaborations. With support from the National Science Foundation INCLUDES program, the MDN focuses on increasing EDI by sharing knowledge, experiences, training, and resources among representatives from multimessenger science collaborations. Representatives to the MDN become engagement leads in their collaboration, extending the reach of the community of practice. An overview of the MDN structure, lessons learned, and how to join are presented.

\vspace{4mm}
{\bfseries Corresponding authors:}
E. Bechtol$^{1*}$, 
K. Bechtol$^{1}$,
S. BenZvi$^{2}$, 
C. Bleve$^{3}$, 
D. Castro$^{4}$, 
B. Cenko$^{5}$, 
L. Corlies$^{6}$, 
A. Furniss$^{7}$, 
C. M. Hui$^{8}$,
D. L. Kaplan$^{9}$, 
J. S. Key$^{10}$, 
J. Madsen$^{1}$, 
F. McNally$^{11}$, 
M. McLaughlin$^{12}$, 
R. Mukherjee$^{13}$, 
R. Ojha$^{5}$, 
J. Sanders$^{14}$, 
M. Santander$^{15}$, 
J. Schlieder$^{5}$, 
D. H. Shoemaker$^{16}$, and
S. Vigeland$^{9}$\\
{$^{1}$ \itshape University of Wisconsin-Madison, WI, USA}
{$^{2}$ \itshape University of Rochester, NY, USA}
{$^{3}$ \itshape Laboratoire de Physique Subatomique et de Cosmologie, Grenoble, France}
{$^{4}$ \itshape Center for Astrophysics, Harvard \& Smithsonian, MA, USA}
{$^{5}$ \itshape NASA Goddard Space Flight Center, MD, USA}
{$^{6}$ \itshape AURA / Vera C. Rubin Observatory, AZ, USA}
{$^{7}$ \itshape California State University East Bay, CA, USA}
{$^{8}$ \itshape NASA Marshall Space Flight Center, AL, USA}
{$^{9}$ \itshape University of Wisconsin-Milwaukee, WI, USA}
{$^{10}$ \itshape University of Washington Bothell, WA, USA}
{$^{11}$ \itshape Mercer University, GA, USA}
{$^{12}$ \itshape West Virginia University, WV, USA}
{$^{13}$ \itshape Barnard College, NY, USA}
{$^{14}$ \itshape Marquette University, WI, USA}
{$^{15}$ \itshape University of Alabama, AL, USA}
{$^{16}$ \itshape MIT Kavli Institute for Astrophysics and Space Research, MA, USA}\\[4mm]
$^*$ Presenter

\FullConference{37$^{\rm{th}}$ International Cosmic Ray Conference (ICRC 2021)\\
		July 12th -- 23rd, 2021\\
		Online -- Berlin, Germany}
		}

\begin{document}
\maketitle

\section{Introduction}
The lack of diversity in physics, astronomy, and astrophysics is well-documented. For example, 2018 data  from the American Institute of Physics' Statistical Research Center shows that only 9\% of physics bachelors degrees were earned by Hispanic American students and 3\% were earned by African American students \cite{aip_stats_aa_ha}, and 78\% of physics bachelor degrees were earned by men while 22\% were earned by women \cite{aip_stats_bachelors}. Increasing diversity in the fields of physics and astronomy is important to ensure scientific progress by making sure all talent is being tapped, in addition to the ethical and social justice motivation. This is especially true in large scientific collaborations which increasingly play a primary role in a researcher's professional interactions and research opportunities. 

The Multimessenger Diversity Network (MDN), formed in late 2018 with four founding members, is now a community of practice comprised of representatives from nine multimessenger astronomy and astrophysics experiments.  Current participating collaborations include the \textit{Fermi} Gamma-ray Space Telescope, IceCube Neutrino Observatory, Laser Interferometer Space Antenna, LIGO Scientific Collaboration, North American Nanohertz Observatory for Gravitational Waves, Pierre Auger Observatory, Neil Gehrels \textit{Swift} Observatory, Vera C. Rubin Observatory, and the Very Energetic Radiation Imaging Telescope Array System. The initial funding was obtained through a supplemental award to IceCube through the NSF INCLUDES (Inclusion across the Nation of Communities of Learners of Underrepresented Discoverers in Engineering and Science) program which aims to transform education and career pathways to broaden participation in science and engineering. As an INCLUDES-funded program, the MDN focuses on broadening participation in multimessenger astronomy by sharing resources, knowledge, experiences, and training among participating collaborations. 

In this proceeding, we offer an overview of the MDN components, highlight initial impacts of the network, share lessons learned, and provide information on how to join the group. 
\begin{figure}
\begin{center}
  \includegraphics[width=4in]{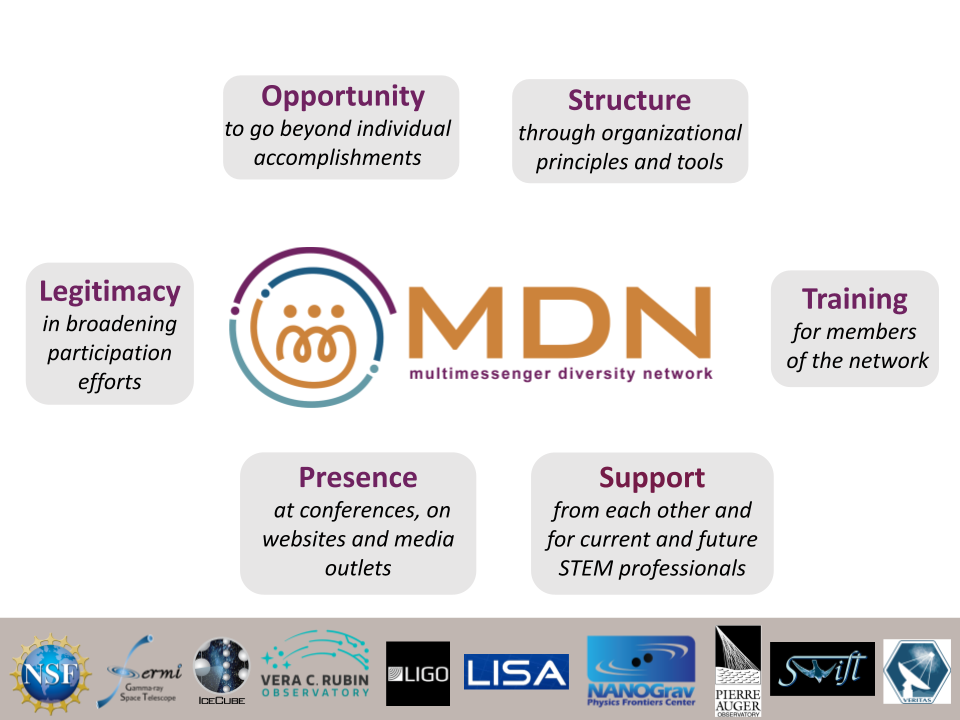}
  \caption{The elements of the MDN and current members, the \textit{Fermi} Gamma-ray Space Telescope, IceCube Collaboration, Vera C. Rubin Observatory, Laser Interferometer Gravitational-Wave Observatory (LIGO), Laser Interferometer Space Antenna, Nanohertz Observatory for Gravitational Waves (NANOGrav), Pierre Auger Observatory, Neil Gehrels \emph{Swift} Observatory, and Very Energetic Radiation Imaging Telescope Array System (VERITAS).}
  \label{fig:MDN}
  \end{center}
\end{figure}

\section{MDN Components}
The MDN is based on a "community of practice" model~\cite{community_practice} introduced to members in 2019 in a community engagement training workshop led by Lou Woodley, Director of the Center for Scientific Collaboration and Community Engagement (\url{www.cscce.org/}). A community of practice is a group of people who care about a subject and carry out activities or share resources on the subject of interest. The MDN relies on and promotes six structural elements shown (Figure~\ref{fig:MDN}) to advance equity, diversity, and inclusion (EDI) in multimessenger collaborations:  (1) \textit{opportunity} to go beyond individual accomplishments, (2) \textit{structure} through organizational principles and tools, (3) \textit{training} for members, (4) \textit{support} from each other and for current and future science, technology, engineering and mathematics (STEM) professionals, (5) \textit{presence} at conferences, on websites, and on media outlets, and (6) \textit{legitimacy} in broadening participation efforts. These elements underpin monthly meetings of the network which provide opportunities for members to share experiences and knowledge, learn from others, and seek and provide support. A website (\url{astromdn.github.io}), along with presentations at conferences, and a submitted Astro2020 State of the Profession white paper ~\cite{mdn_whitepaper}, lend legitimacy to the often volunteer-based EDI efforts of collaborations. 

A part-time community manager runs the monthly meetings, responds to inquiries, helps represent the MDN at conferences, and promotes engagement within the group. This position has proven critical to the sustainability and longevity of the network. The community manager keeps the group connected, engaged, and helps rally members around a few key activities that catalyze interactions. In the past, activities included co-authoring an Astro2020 State of the Profession white paper, participating in a two-day community engagement training, and hosting guest speakers. Some planned activities were paused due to the COVID-19 pandemic. 

\begin{figure}
\begin{center}
  \includegraphics[width=4in]{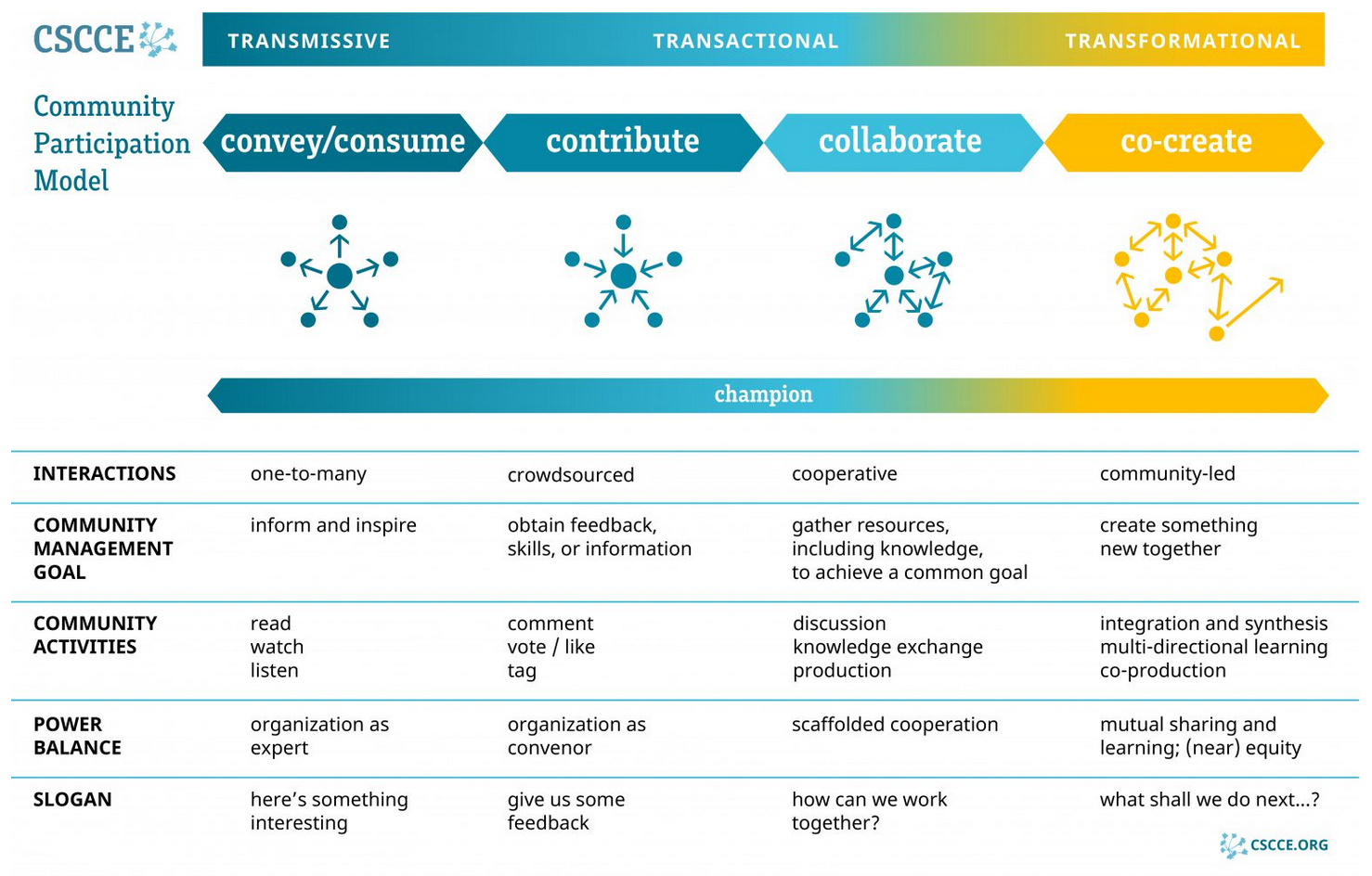}
  \caption{The Community Participation Model from the Center for Scientific Collaboration and Community Engagement~\cite{cscce_community_model} describes four modes of community member participation. With both cooperative and community-led interactions, participation in the MDN falls under "collaborate" and "co-create" community models.}
  \label{fig:CSCCE}
  \end{center}
\end{figure}

 As a community of practice, the MDN focuses on EDI in multimessenger astronomy collaborations with regular activities including monthly meetings, maintaining a website, and participating in conferences. The Community Participation Model from the Center for Scientific Collaboration and Community Engagement, Figure \ref{fig:CSCCE}, sheds light on stages of community development and has been a useful model for the MDN. As Woodley and Pratt \cite{cscce_community_model} write, communities often start in the "convey/consume" phase with information moving out from a community manager to community members, and move towards a "co-create" phase where members work together to create something new. Reflecting on the MDN as a community, it has occupied each community participation phase at some point, with much time spent advancing and retreating between "collaborate" and "co-create." Referencing Figure \ref{fig:CSCCE}, the network's interactions, goals, and activities are most aligned with these two phases of community participation.  
%https://www.cscce.org/resources/community-participation-model/

%Monthly meetings serve as a space for people to share-out and learn from each other as well as guest speakers. During a typical call, there is dedicated time for updates and current happenings in collaborations. This is often a time when 
%If need more can add about community state (i.e. collaborate and co-create; https://www.cscce.org/resources/community-participation-model/). 

\section{Initial Impacts}
The strength of the MDN lies in the community connections which provide opportunities to share experiences,  describe lessons learned, exchange documentation for best practices such as a code of conduct, and present models of a variety of EDI efforts.  Raising awareness of EDI work among members is one positive outcome on the MDN effort. Here we highlight three examples exemplifying the community of practice model where members are co-creators and freely exchange ideas and materials for others to adapt and adopt. 

The VERITAS Collaboration has moved forward with two major DEI efforts, namely the recognition of collaboration service through annual VERITAS Outstanding Contribution Awards, as well as the adoption of an official Collaboration Code of Conduct.  Both of these positive changes were motivated through the communication of procedures already in place within other MDN member collaborations. In the case of the service awards, which are given to early-career scientists in recognition of critical contributions to the collaboration, the IceCube Impact Awards were a model. The VERITAS collaboration Code of Conduct was motivated by and styled after the \textit{Fermi}-LAT collaboration version.  

In bringing together several multimessenger collaborations, it is clear that many groups are choosing similar actions to address their EDI disparities. Two topics of interest within the MDN over the last two and half years are collaboration-wide surveys and mentoring programs. 

(1) Collaboration-wide surveys: In the course of pursuing EDI initiatives, two questions that often come up are, "How will we know if our efforts are working if we don't have any demographic information about our members?" and "How does the collaboration experience differ by group (broken out by gender, career level, and/or race)?" One way to address one or both of these questions is a collaboration-wide survey. Such a survey could focus on collecting demographic data or focus on the climate, or atmosphere, of a collaboration. As an example within the MDN, the survey experiences of NANOGrav have directly informed the discussions of a survey within the IceCube collaboration. While the IceCube survey effort is in the very early planning phases, it has been invaluable to hear how another geographically distributed collaboration has developed and evolved their own survey strategy. NANOGrav representatives have shared with some frequency about their collaboration-wide surveys, not just about the planning and execution, but critically about how the collaboration takes action based on the survey results. As a result of recent surveys, NANOGrav has put in place more support structures for post-doctoral scholars and graduate students, and restructured its membership to be more transparent and accessible to early career researchers. Such examples are a reminder that running a survey should not be an end goal.  They are one way to collect data on the pulse of a collaboration and help set a course for meaningful and sustainable change. Sharing the lessons from NANOGrav with the other eight collaborations shows the value of the MDN to leverage the effort of one group to benefit a much larger number of researchers and scholars.

(2) Mentoring programs: In the case of mentorship, a recent experience highlights the potential of having a community of practice focused on EDI in mutlimessenger astronomy. The suggestion for a meeting on mentorship came from the \textit{Fermi}-LAT representative, who connected the MDN community manager with the mentoring program lead. Members of the MDN were encouraged to invite collaborators to join and several new individuals were welcomed on the call. After a presentation on the \textit{Fermi}-LAT graduate student mentoring program, the group asked questions and engaged in discussion. The \textit{Fermi}-LAT group graciously shared their resources with the MDN including a program proposal and interest forms,  and encouraged others to use them as springboards for mentoring programs in their own collaborations. This had a direct impact on the LISA Consortium, which is now pursuing a mentoring program. This generous sharing of knowledge and experience is at the core of the MDN and amplifies the work of a few across the field at-large. 

%As on of NSF's 10 Big Ideas, the approach of multimessenger astrophysics has enjoyed increased and renewed funding and recognition. With this comes growth in multimessenger collaborations and an increase in the amount of time scientists, researchers, and scholars spend interacting with individuals from around the world. At the same time, there is growing recognition in the value of diverse perspectives and the importance of graduate students and early career scientists having ways to communicate

\section{Lessons Learned}
Since the MDN was formed in 2018, EDI in STEM have taken on a new level of recognition and importance. Now more than ever, researchers, scientists, and scholars are engaged in work to make STEM a more equitable, diverse, inclusive, and accessible field. Here, we share lessons learned from the last two and half years.

Despite the growth in awareness and importance of EDI in STEM, this work is still overwhelmingly volunteer-based, i.e. not necessarily counted in the work load. It is challenging for individuals, many of whom were already addressing work-life balance issues and are often from underrepresented groups, to dedicate time to EDI efforts. This reality has implications for how people participate in the MDN. For instance, a regular meeting cadence is sometimes the only opportunity for engagement during the month and has become the key vehicle for knowledge sharing and community building. Additionally, planned activities have been adapted or abandoned based on the interest and availability of members. At the start of the MDN, it was thought that members would develop a "playbook"~\cite{playbook} or "guidebook" of EDI in their collaboration. The idea and documentation were proposed and included in a training event. Although members recognized that such a guidebook could be useful, few had the time or energy to invest in its development. The network pivoted to rely on knowledge sharing \textit{among} collaborations during the monthly meetings and through a dedicated Slack organization. For now, this level of engagement is sufficient to support members as they work for organizational change with the time and energy they have available. We continue to push for a recognition of EDI work as critical for the well-being and advancement of scientific collaborations. Counting EDI work towards collaboration-level service, highlighting EDI work in press releases, and developing EDI-focused leadership roles are actionable steps collaborations could take to recognize this important effort ~\cite{mdn_whitepaper}.

\section{Joining MDN}
When the MDN formed, it was unclear how much interest there would be in such a community. Through contacts and professional networks, the MDN grew quickly from four to six to nine member groups; there are regular expressions of interest from the broader fields of high-energy, astroparticle physics, and cosmology. If you are interested in joining or following our work, please visit \url{astromdn.github.io} and subscribe to our Google Group, or write directly to the contact authors. We welcome your participation and interest. 

\section{Summary and Outlook}
 STEM professionals continue to be overwhelmingly male and do not reflect the overall racial and ethic makeup of the world as a whole.  We described the MDN, a community of practice with representatives from nine larger multimessenger astronomy and astrophysics experiments, which is working to increase awareness of EDI issues and achieve better outcomes in the multimessenger community. The elements of the MDN and its structure were described as well as the central role of the community manager who acts as a facilitator to promote interactions and exchanges within the network.  A few of the lessons learned by the MDN were introduced and an invitation with instructions on how to get more information was presented.  
 
 The outlook for the MDN is encouraging, with a core group that meets virtually monthly in a supportive, educational environment.  The EDI challenges are significant and, as described, often being addressed on top of other expectations that usually have higher priority. The MDN promotes the value of EDI to both achieve better outcomes and also the recognition of the need and value of this work.

\clearpage
\clearpage

\section*{Full Author List: IceCube Collaboration}

% \noindent \textbf{Note comment afterwards:} Collaborations have the possibility to provide an authors list in xml format which will be used while generating the DOI entries making the full authors list searchable in databases like Inspire HEP. For instructions please go to icrc2021.desy.de/proceedings or contact us under icrc2021proc@desy.de.\\

% \scriptsize
% \noindent
% first.author$^1$, 
% second.author$^2$, 
% third.author$^3$ % .... more names
% and 
% last.author$^{n}$ \\

% \noindent
% $^1$first.affiliation.
% $^2$second.affiliation. % .... more affiliation
% $^{m}$last.affiliation.

\scriptsize
\noindent
R. Abbasi$^{17}$,
M. Ackermann$^{59}$,
J. Adams$^{18}$,
J. A. Aguilar$^{12}$,
M. Ahlers$^{22}$,
M. Ahrens$^{50}$,
C. Alispach$^{28}$,
A. A. Alves Jr.$^{31}$,
N. M. Amin$^{42}$,
R. An$^{14}$
% ...
\\

\noindent
% ...
$^{12}$ Universit{\'e} Libre de Bruxelles, Science Faculty CP230, B-1050 Brussels, Belgium \\
% ...
$^{14}$ Department of Physics and Laboratory for Particle Physics and Cosmology, Harvard University, Cambridge, MA 02138, USA \\
% ...
$^{17}$ Department of Physics, Loyola University Chicago, Chicago, IL 60660, USA \\
$^{18}$ Dept. of Physics and Astronomy, University of Canterbury, Private Bag 4800, Christchurch, New Zealand \\
% ...
$^{22}$ Niels Bohr Institute, University of Copenhagen, DK-2100 Copenhagen, Denmark \\
% ...
$^{28}$ D{\'e}partement de physique nucl{\'e}aire et corpusculaire, Universit{\'e} de Gen{\`e}ve, CH-1211 Gen{\`e}ve, Switzerland \\
% ...
$^{31}$ Karlsruhe Institute of Technology, Institute for Astroparticle Physics, D-76021 Karlsruhe, Germany  \\
% ...
$^{42}$ Bartol Research Institute and Dept. of Physics and Astronomy, University of Delaware, Newark, DE 19716, USA \\
% ...
$^{50}$ Oskar Klein Centre and Dept. of Physics, Stockholm University, SE-10691 Stockholm, Sweden \\
% ...
$^{59}$ DESY, D-15738 Zeuthen, Germany \\
% ...

\subsection*{Acknowledgements}

\noindent
USA {\textendash} U.S. National Science Foundation-Office of Polar Programs,
U.S. National Science Foundation-Physics Division,
U.S. National Science Foundation-EPSCoR,
Wisconsin Alumni Research Foundation,
Center for High Throughput Computing (CHTC) at the University of Wisconsin{\textendash}Madison,
Open Science Grid (OSG),
Extreme Science and Engineering Discovery Environment (XSEDE),
Frontera computing project at the Texas Advanced Computing Center,
U.S. Department of Energy-National Energy Research Scientific Computing Center,
Particle astrophysics research computing center at the University of Maryland,
Institute for Cyber-Enabled Research at Michigan State University,
and Astroparticle physics computational facility at Marquette University;
Belgium {\textendash} Funds for Scientific Research (FRS-FNRS and FWO),
FWO Odysseus and Big Science programmes,
and Belgian Federal Science Policy Office (Belspo);
Germany {\textendash} Bundesministerium f{\"u}r Bildung und Forschung (BMBF),
Deutsche Forschungsgemeinschaft (DFG),
Helmholtz Alliance for Astroparticle Physics (HAP),
Initiative and Networking Fund of the Helmholtz Association,
Deutsches Elektronen Synchrotron (DESY),
and High Performance Computing cluster of the RWTH Aachen;
Sweden {\textendash} Swedish Research Council,
Swedish Polar Research Secretariat,
Swedish National Infrastructure for Computing (SNIC),
and Knut and Alice Wallenberg Foundation;
Australia {\textendash} Australian Research Council;
Canada {\textendash} Natural Sciences and Engineering Research Council of Canada,
Calcul Qu{\'e}bec, Compute Ontario, Canada Foundation for Innovation, WestGrid, and Compute Canada;
Denmark {\textendash} Villum Fonden and Carlsberg Foundation;
New Zealand {\textendash} Marsden Fund;
Japan {\textendash} Japan Society for Promotion of Science (JSPS)
and Institute for Global Prominent Research (IGPR) of Chiba University;
Korea {\textendash} National Research Foundation of Korea (NRF);
Switzerland {\textendash} Swiss National Science Foundation (SNSF);
United Kingdom {\textendash} Department of Physics, University of Oxford.

\end{document}